\documentstyle[psfig]{elsart}

\def\lsim{\mathrel{\rlap{\lower4pt\hbox{\hskip1pt$\mathchar"0218$}}
       \raise1pt\hbox{$<$}}}
\def\gsim{\mathrel{\rlap{\lower4pt\hbox{\hskip1pt$\mathchar"0218$}}
       \raise1pt\hbox{$>$}}}

\begin{document}
\begin{frontmatter}
\title{\vspace*{-2cm}{\small\parskip=0pt\baselineskip=12pt\hspace*{1in}
For proceedings of {\it Frontiers in Neutron Scattering Research},\\
\hspace{2in} ISSP, University of Tokyo, Nov.\ 24--27, 1998.\\ 
\vspace{-9pt}\hspace{2.2in} To be published in {\it J. Phys. Chem. Solids}}\\
\vspace*{1.2cm} CHARGE SEGREGATION AND ANTIFERROMAGNETISM\\ IN HIGH-$T_c$
SUPERCONDUCTORS}
\author{J. M. Tranquada,$^\dagger$ N. Ichikawa,$^\ddagger$ K.
Kakurai,$^{\mbox{\protect\scriptsize \S}}$ and S. Uchida$^\ddagger$}
\address{$^\dagger$Physics Department, Brookhaven National Laboratory, Upton, NY
11973, USA}
\address{$^\ddagger$Department of Applied Physics, The University of Tokyo,
Yayoi 2-11-16, Bunkyo-ku, Tokyo 113, Japan}
\address{$^{\mbox{\protect\scriptsize\S}}$Institute for Solid State Physics, The
University of Tokyo, Roppongi, Minato-ku, Tokyo 106-8666, Japan}
\begin{abstract}
Local antiferromagnetism coexists with superconductivity in the cuprates.  Charge
segregation provides a way to reconcile these properties.  Direct evidence for
modulated spin and charge densities has been found in neutron and X-ray
scattering studies of Nd-doped La$_{2-x}$Sr$_x$CuO$_4$.  Here we discuss the
nature of the modulation, and present some new results for a Zn-doped sample. 
Some of the open questions concerning the connections between segregation and
superconductivity are described.
\end{abstract}
\end{frontmatter}


\section{INTRODUCTION}

Neutron-scattering, nuclear-magnetic-resonance (NMR), and muon-spin-ro\-ta\-tion
($\mu$SR) studies have all provided evidence for the coexistence of local
antiferromagnetic spin correlations with superconductivity in the layered
cuprates.  The nature of this coexistence has been the subject of considerable
debate.  Neutron scattering studies of certain cuprate superconductors
\cite{hayd96a,bour97} have indicated that the magnetic excitations are
surprisingly similar to those in the undoped parent compounds, which are both
insulating and antiferromagnetic due to strong electronic interactions.  One
way to accomodate both antiferromagnetic correlations and mobile holes is
through charge segregation. 

One type of segregation of particular interest involves periodically-spaced
charge stripes that act as antiphase domain walls between narrow
antiferromagnetic domains.  Perhaps the first evidence of such correlations was
the measurement of inelastic magnetic scattering peaked at incommensurate wave
vectors in La$_{2-x}$Sr$_{x}$CuO$_4$ with $x\approx0.15$
\cite{cheo91,maso92,thur92}.  Study of the charge modulation has become
possible with the discovery that, for $x\sim\frac18$, stripes can be pinned by
the structural modulation induced by partial substitution of the smaller ion 
Nd$^{3+}$ for La$^{3+}$ \cite{tran95a,tran96b}.  The charge order originally
detected by neutron scattering has been confirmed with high-energy x-rays
\cite{vonz98}, and there is considerable evidence for intimate coexistence of
stripe order and superconductivity at $x=0.15$ \cite{tran97a,oste97,nach98}.

Elastic incommensurate magnetic peaks have also been discovered in Zn-doped 
La$_{2-x}$Sr$_{x}$CuO$_4$ with $x=0.14$ \cite{hiro98}, and even in samples with
no Zn and $x=0.12$ \cite{suzu98,kimu98}.  In samples of
La$_{2-x}$Sr$_{x}$CuO$_4$ with no static order, 
incommensurate splitting of the inelastic magnetic scattering is nearly
identical to that of the elastic peaks in Nd-doped samples with the same Sr
concentration \cite{yama98a}.  The possibility that the magnetic scattering in
underdoped YBa$_2$Cu$_3$O$_{6+x}$ is incommensurate had been considered earlier
\cite{ster94,tran97d}, and recently it has been demonstrated that, indeed, at
least part of the scattering is incommensurate \cite{dai98}, with 
modulation wave vectors consistent with the 214 system \cite{mook98a}.

In this paper, we briefly discuss several topics related to stripes in the
cuprates.  In the next section we review the connection between the
experimentally observed superlattice peaks, and the real-space modulations of
the spin and charge densities.  In section 3, we present some new results on a
Zn-doped 214 sample.  Finally, in section 4 we list some of the open questions
concerning stripes and superconductivity in the cuprates.

\section{NATURE OF THE SPIN AND CHARGE DENSITY MODULATIONS}

There have been questions raised concerning the interpretation of the
observed superlattice peaks that suggest some confusion over the distinction
between scattering from a 1D system and that from a 2 or 3D system with a 1D
modulation.  In order to clarify things, first consider a line of atoms
with a spacing $a$.  In reciprocal space, the scattering from such a 1D system
consists of constant-intensity sheets separated by $2\pi/a$.

In contrast, consider a Bravais lattice of atoms with positions ${\bf R}_j$, and
suppose that the positions undergo a small sinusoidal modulation of the form
\begin{equation}
  {\bf u}_j = {\bf A}\sin({\bf g}\cdot {\bf R}_j+\phi),
\end{equation}
where {\bf A} is the amplitude, {\bf g} is the modulation wave vector, and $\phi$
is an arbitrary phase shift that we will ignore.  Overhauser \cite{over71} has
shown that the scattering from such a system is given by
\begin{equation}
  I({\bf Q}) = \sum_{{\bf G},n} J_n^2({\bf Q}\cdot{\bf A})
   \delta({\bf Q}-{\bf G}-n{\bf g}).
\end{equation}
When $n=0$ the scattering corresponds to fundamental Bragg peaks with 
\begin{equation}
  I({\bf G}) \approx 1-\frac12({\bf Q}\cdot{\bf A})^2 \approx
    e^{-({\bf Q}\cdot{\bf A})^2/2},
\end{equation}
for ${\bf Q}\cdot{\bf A}\ll 1$.  The modulation causes an intensity reduction
with a form similar to a Debye-Waller factor. For
$n=1$, one finds superlattice peaks split about each reciprocal lattice vector
{\bf G} by {\bf g}, with
\begin{equation}
I({\bf G}+{\bf g}) \approx \frac14({\bf Q}\cdot{\bf A})^2.
\end{equation}
Higher-order peaks will also occur, but they are extremely weak for small 
${\bf Q}\cdot{\bf A}$.  For example, 
$I({\bf G}+2{\bf g})\approx\frac14 I^2({\bf G}+{\bf g})$.

The configuration of superlattice peaks observed in stripe-ordered\linebreak
La$_{1.6-x}$Nd$_{0.4}$Sr$_x$CuO$_4$ has been reviewed elsewhere \cite{tran98b}. 
Briefly, there are two sets of superlattice peaks, which are most easily
described in terms of a unit cell with $a\approx3.8$~\AA.  One set of peaks is
split about the antiferromagnetic wave vector by an amount
$\epsilon\times2\pi/a$ along the [100] and [010] directions, indicating
antiphase antiferromagnetic domains.  A second set of peaks occurs about
nuclear Bragg peaks, split by $2\epsilon\times2\pi/a$, indicative of
charge-order.  Given that we have peaks split in two directions, there are two
possible interpretations: either we are averaging over domains each of which
has a single modulation direction, or the two modulations are superimposed in
each region of the sample \cite{tran98a}.  The simple stripe picture is based
on the former model.  Can we rule out the latter?

If there is a stripe grid, then the phase of the antiferromagnetic domains must
be modulated in two directions.  The arrangement of domains and their relative
phase factors forms a checker-board structure, similar to a simple N\'eel
antiferromagnet.  The unit cell of such a structure has its axes rotated by
$\pi/4$ relative to the modulation directions, and it has an area twice that of
a single domain.  This means that in reciprocal space, the first superlattice
peaks should be rotated by $\pi/4$ relative to the modulation wave vectors. 
Experimentally, we have checked for magnetic peaks split along [110] and
$[1\bar{1}0]$, and have found nothing.  (In principle, a stripe grid should
result in a square lattice of superlattice peaks, so that there should be
peaks in these directions regardless of the modulation directions; however, they
might be anomalously weak.)  Hence, using the grid interpretation, the peaks
split along [100] and [010] imply stripe modulations in real space along [110]
and $[1\bar{1}0]$ (i.e., diagonal stripes).

Unlike the magnetic peaks, the charge-order peaks from such a grid of diagonal
stripes should not be rotated.  The first charge order peaks should appear at
$(\epsilon,\epsilon,0)$, so that the peaks we have observed at
$(0,2\epsilon,0)$ would involve the sum of the two modulation wave vectors.  To
test this possibility, we performed neutron scattering measurements on the
PONTA (5G) triple-axis spectrometer at the JRR-3 reactor at JAERI. 
The $x=0.12$ Nd-doped crystal characterized previously \cite{tran96b} was
used, and elastic scans in the appropriate directions were performed (using
14.7-meV neutrons and relatively open collimation) at two temperatures: 7~K and
65~K.  We subtracted the data at 65~K from the 7~K measurement in order to
isolate any signal that might appear only at low temperature.  The results are
shown in Fig.~1.  The peak found in the [010] direction is consistent with
previous work \cite{tran96b}; however, only noise is present along [110]. 
Thus, a 2D grid interpretation appears to be incompatible with experiment.

\begin{figure}
\centerline{\psfig{figure=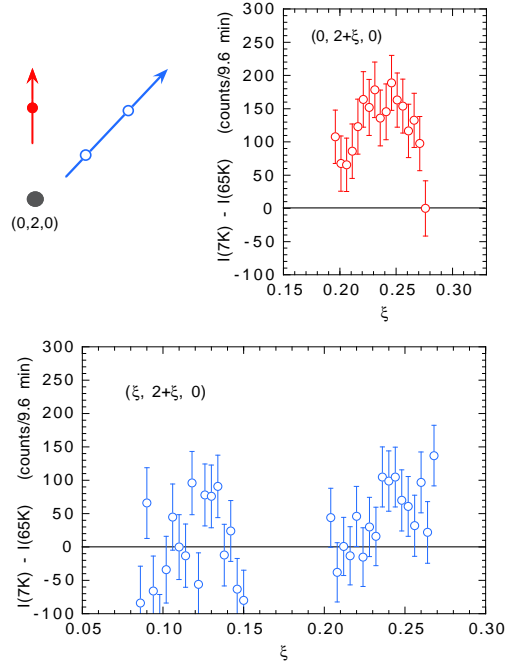,width=7cm}}
\caption{Search for charge-order peaks in
La$_{1.48}$Nd$_{0.4}$Sr$_{0.12}$CuO$_4$.  Upper left indicates scan directions
in reciprocal space.  Upper right: intensity difference between 7~K and 65~K
scans along [010] direction.  Lower: similar measurement along [110].}
\end{figure}

\section{STATIC MODULATIONS IN Zn-DOPED La$_{2-x}$Sr$_x$CuO$_4$}

It is of interest to search for evidence of charge-stripe order in other
cuprate systems.  So far, charge-order superlattice peaks have only been
observed in Nd-doped La$_{2-x}$Sr$_x$CuO$_4$, where the stripes are pinned by
the low-temperature-tetragonal lattice structure.  One obvious candidate system
is Zn-doped\linebreak 
La$_{2-x}$Sr$_x$CuO$_4$.  Elastic magnetic peaks have been
observed by Hirota {\it et al.} \cite{hiro98} in a superconducting crystal with
$x=0.14$.

With the intention of looking for charge order, a single crystals
($\approx0.45$~cm$^3$) of La$_{1.88}$Sr$_{0.12}$Cu$_{0.98}$Zn$_{0.02}$O$_4$ was
grown with an infrared image furnace at the University of Tokyo.  Again, we made
use of the PONTA (5G) triple-axis spectrometer, with an incident energy of
14.7~meV and a pyrolytic graphite filter before the sample.  Relatively tight
collimation was used to measure the lattice parameters, yielding $a=5.2657$~\AA\
and $b=5.3042$~\AA\ ($b-a=0.0385$~\AA) at 40~K.  Although the crystal is
orthorhombic at low temperature, we chose to work in tetragonal coordinates
($a\approx b=3.74$~\AA) to search for elastic magnetic peaks.  Opening the
horizontal collimations to
$40'$-$40'$-$80'$-$80'$, we scanned along ${\bf Q}=(\frac12,\frac12+\xi,0)$ and
found peaks at $\xi=\pm0.121\equiv\pm\epsilon$.  An example is shown in
Fig.~2.  The peak width is roughly 40\%\ greater than that found under the same
conditions in La$_{1.48}$Nd$_{0.4}$Sr$_{0.12}$CuO$_4$, with no correction for
resolution.  The temperature dependence of the peak intensity is presented in
Fig.~3.  The disordering temperature of $\sim20$~K is intermediate with respect
to those ($T_m\sim30$~K and 17~K) found in crystals of
La$_{1.88}$Sr$_{0.12}$Cu$_{1-y}$Zn$_y$O$_4$ with $y=0$ and 0.03 (respectively)
by Kimura {\it et al.} \cite{kimu98}.  Initial attempts to observe charge-order
peaks were unsuccessful. 

\begin{figure}
\centerline{\psfig{figure=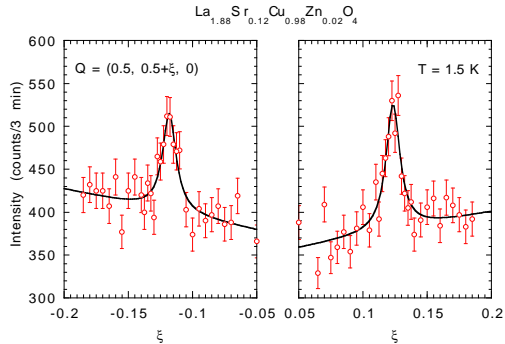,width=7cm}}
\caption{Elastic scans through magnetic peaks at $T=1.5$~K in
La$_{1.88}$Sr$_{0.12}$Cu$_{0.98}$Zn$_{0.02}$O$_4$.}
\end{figure}

\begin{figure}
\centerline{\psfig{figure=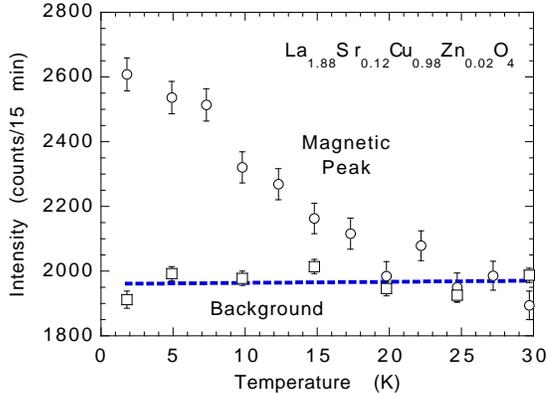,width=7.4cm}}
\caption{Temperature dependence of magnetic peak intensity in\protect\linebreak
La$_{1.88}$Sr$_{0.12}$Cu$_{0.98}$Zn$_{0.02}$O$_4$.  The dashed line is a linear
fit to the background measurements.}
\end{figure}

$\mu$SR \cite{nach96} and NMR \cite{maha94} studies indicate that Zn
suppresses superconductivity locally, resulting in electronic inhomogeneity. 
One might expect that if Zn serves to pin stripes, this effect would also be
inhomogeneous.  Hence, it is of interest to compare the magnetic peak intensity
with that found in La$_{1.48}$Nd$_{0.4}$Sr$_{0.12}$CuO$_4$, where $\mu$SR has
shown the magnetic order to be relatively uniform \cite{nach98}.  We measured
the same magnetic peaks in the latter compound under identical conditions,
except that we worked at a temperature of 7~K in order to avoid any significant
contribution from the Nd moments.  The relative crystal volumes were determined
by phonon measurements.  Normalizing by volume, and assuming no substantial
difference in $l$ dependence of the scattering, the magnetic intensity in the
Zn-doped crystal was found to be just $(22\pm6)\%$ of that in the Nd-doped
crystal.  If this represents the volume fraction showing stripe order, then it
is not surprising that charge-order peaks were not observed.

\section{OPEN QUESTIONS}

While there is substantial evidence for stripe correlations in
La$_{2-x}$Sr$_x$CuO$_4$, the issue of whether charge stripes are common to all
superconducting cuprates remains controversial.  The recent observations
\cite{dai98,mook98a} of incommensurate magnetic scattering in
YBa$_2$Cu$_3$O$_{6.6}$ provide an important connection.  One would also like to
see evidence of related charge correlations, but this is more difficult.  Mook
and coworkers have made some progress in this direction using a special
energy-integrated technique \cite{mook98b}.  It may also be necessary to
investigate phonon anomalies, such as the high-energy longitudinal optical 
branch in La$_{1.85}$Sr$_{0.15}$CuO$_4$ that has been studied by Egami and
coworkers \cite{mcqu98}.

Another issue in YBa$_2$Cu$_3$O$_{6+x}$ concerns the so-called resonance peak. 
The resonance peak is centered on the antiferromagnetic wave vector, and the
energy at which it is centered increases with $x$.  Bourges \cite{bour98} has
noted that the ratio of the resonance-peak energy to the superconducting
transition temperature, $T_c$, is roughly constant.  In the BCS model for
superconductivity, the ratio of the superconducting gap to $T_c$ is also a
constant.  This similarity might lead one to suspect a connection between the
resonance-peak energy and the superconducting gap.  However, measurements of
the gap in Bi$_2$Sr$_2$CaCu$_2$O$_{8+\delta}$ by
tunneling \cite{oda97,renn98,miya98} and photoemission \cite{harr96,ding96}
indicate that the superconducting gap increases while $T_c$ decreases on the
underdoped side.  While similar measurements are not yet available for
YBa$_2$Cu$_3$O$_{6+x}$, infrared conductivity studies of the latter system
suggest that the size of the gap does not decrease as $x$ is reduced from 1
\cite{baso96}.  What is the possible relationship between the resonance peak
and stripe correlations?

So far, we have discussed only hole-doped superconductors.  Might stripes be
relevant to electron-doped superconductors?  Moving beyond cuprates, charge
stripes are already known to be important in nickelates and certain
manganates.  Do stripes occur in other transition-metal oxide systems? 
Clearly, there is a great deal of work left to do, and neutron scattering will
be a prominent tool in this effort.

\section{ACKNOWLEDGMENT}

Work at Brookhaven is supported by Contract No.\ DE-AC02-98CH10886, Division
of Materials Sciences, U.S. Department of Energy.


\end{document}